\documentclass[aps,pra,twocolumn,showpacs,showkeys,amsmath,amssymb]{revtex4-1}

\def\Eq#1{Eq.~(\ref{#1})}
\def\Eqs#1{Eqs.~(\ref{#1})}

\def\no{\nonumber\\}
\def\r{\rangle\!\rangle}
\def\l{\langle\!\langle}
\def\>{\rangle}
\def\<{\langle}

\def\half{\tfrac{1}{2}}
\def\U{U}
\def\adg{a^\dagger}
\def\rhoh{\hat{\rho}}
\def\rhot{\tilde{\rho}}

\def\dg{\dagger}
\def\A{A}
\def\At{\widetilde{A}}
\def\Aa{\widetilde{A}}
\def\lam{\lambda}

\def\ome{\omega}
\def\del{\delta}
\def\sig{\sigma}
\def\gam{\gamma}

\def\al{\alpha}
\def\bt{\beta}

\def\Hm{\mathcal{H}}
\def\Lm{\mathcal{L}}
\def\L{L}
\def\xth{\hat{\widetilde{x}}}
\def\xh{\hat{x}}
\def\ph{\hat{p}}
\def\pth{\hat{\widetilde{p}}}
\def\xt{\widetilde{x}}

\def\d{\partial}

\def\Xh{\hat{X}}

\begin{document}

\title{Thermal symmetry of the Markovian master equation}

\author{B. A. Tay }
\altaffiliation{Also at: Institute for Mathematical Research,
University Putra Malaysia, 43400 UPM Serdang, Selangor, Malaysia.}
\email{batay@science.upm.edu.my} \affiliation{Department of Physics,
Faculty of Science, University Putra Malaysia, 43400 UPM Serdang,
Selangor, Malaysia.}

\author{T. Petrosky}\email{petrosky@physics.utexas.edu}
\affiliation{Center for Complex Quantum Systems, The University of
Texas at Austin, 1 University Station C1600, Austin, TX 78712, USA.}

\date{\today}

\begin{abstract}

The quantum Markovian master equation of the reduced dynamics of a
harmonic oscillator coupled to a thermal reservoir is shown to
possess thermal symmetry. This symmetry is revealed by a Bogoliubov
transformation that can be represented by a hyperbolic rotation
acting on the Liouville space of the reduced dynamics. The Liouville
space is obtained as an extension of the Hilbert space through
the introduction of tilde variables used in the thermofield dynamics formalism.
The angle of rotation depends on the temperature of the reservoir,
as well as the value of Planck's constant. This symmetry relates the thermal
states of the system at any two temperatures. This includes absolute
zero, at which purely quantum effects are revealed. The Caldeira-Leggett
equation and the classical Fokker-Planck equation also possess
thermal symmetry. We compare the thermal symmetry obtained from the
Bogoliubov transformation in related fields and discuss the effects
of the symmetry on the shape of a Gaussian wave packet.

\end{abstract}


\maketitle

\section{Introduction}

In the dynamical formulation of a quantum system in the Hilbert
space $\Hm$, a continuous symmetry is represented by a unitary
transformation $U$ that commutes with the Hamiltonian $H$
\cite{Wigner}. When we extend the dynamics to the Liouville space
$\Lm= \Hm \times \tilde{\Hm}$, the symmetry of $H$ is translated
into a symmetry of the Liouville operator, defined by $L=-i(H \times
1 - 1 \times H)$ for a system with unitary time evolution. However,
the reverse is not true. The Liouville operator may contain a
symmetry that has no counterpart in the Hamiltonian \cite{Brink01}.
For example, consider a simple harmonic oscillator with Hamiltonian
$H_0=\ome_0 a^\dg a$. This system possesses a non-degenerate
discrete energy spectrum, $E_n$. Nevertheless, the energy spectrum
of the corresponding Liouville operator $L_0=-i(H_0 \times 1 - 1
\times H_0 )$ depends on energy differences $E_m-E_n$, and therefore
degeneracies may occur \cite{Petrosky97}. The occurrence of
degeneracies in the spectrum of the Liouville operator is a
signature for the presence of a symmetry in $\Lm$ that has no
counterpart in $\Hm$.

It often occurs that the system we are interested in is embedded in
a larger system, for instance a thermal reservoir. The subsystem is
an open interacting quantum system with a non-unitary (semigroup)
time evolution. We obtain the reduced dynamics of the system by
tracing out the degrees of freedom of the reservoir. When memory
effects are further neglected, we obtain the quantum Markovian
master equation (MME) \cite{Kossa}. This formalism provides a
framework for describing dissipative processes in various fields,
such as quantum optics \cite{Weidlich65,Agarwal}, quantum Brownian
motion \cite{Weiss,Gardiner}, quantum decoherence \cite{Zurek03} and,
more recently, quantum information theory \cite{Nielson}.  The
generator of the time evolution of the MME is a non-Hermitian
collision operator with complex eigenvalues \cite{Briegel93,Tay04}.
For different thermal states of the system, degeneracies may occur
in the eigenvalues of the collision (or Liouville) operator. These degeneracies
indicate the presence of a {\it thermal} symmetry that is the main
subject of this paper. In contrast to the discussion in
Ref.~\cite{Brink01}, in which the authors considered an exact symmetry $S$ of the
Liouville operator in the sense of $[S,L]=0$, we consider a symmetry
$U$ of the collision operator $K(b)$ up to a similarity
transformation, $K(b')=U K(b) U^\dg$, where $b$ labels the different
thermal states of the system.

A natural setting for formulating our discussion is provided by the
thermofield dynamics (TFD) formalism \cite{Umezawa}. In this
formalism, the extension to the Liouville space is achieved by
introducing the operator $A=a \times 1$ and the tilde conjugate
operator $ \At^\dg = 1 \times a$ for the annihilation operator $a$
of the harmonic oscillator. A Bogoliubov transformation, which can
be represented by a hyperbolic rotation, can then be defined to act
on the pair of operators $(A,\At^\dg)$ and its Hermitian conjugate
pair \cite{Santana06}. Thermal effects of the reservoir enter our
analysis through a temperature dependent angle in the hyperbolic
rotation. As a result, the rotation connects different thermal
states of the system; in particular, each temperature is related to
the zero temperature that contains purely quantum effects (without
thermal effects).

A thermal symmetry may also exist in a classical system, though it
occurs in a different form. This is true for the classical
Fokker-Planck equation \cite{FokkerPlanck} where the thermal
symmetry results from a scale transformation on the phase space
variables. We find that the Caldeira-Leggett equation \cite{CL}
possesses a thermal symmetry as well. However, when we consider the
thermal symmetry of MMEs derived under more general conditions
\cite{Unruh89,HPZ92,Tay06}, the thermal symmetry is no longer
present.

Since the Bogoliubov transformation preserves the canonical
commutation relations between the position and momentum operators, we can
introduce a symplectic structure into the theory. The thermal
symmetry can then be formulated as an element of the
$\text{Sp}(2n,\mathbb{R})$ group acting on the phase space operators, which
is often employed in the studies of ``Gaussons," or Gaussian pure states
\cite{Sudarshan87}. We also discuss how the generator of the thermal
symmetry is related to the Lie algebra of the $\text{SU}(1,1)$ group utilized in quantum optics \cite{Ban93} and compare the thermal symmetry to other canonical transformation discussed in the literature \cite{Ekert90}.

Since the thermal symmetry connects different thermal states of the system, it also dictates the change in shape of the density function in the coordinate space as a function of the temperature. It is known that a density function with Gaussian profile experiences simultaneous stretch and contraction along the diagonal and off-diagonal directions separately in the coordinate space as a result of a change in temperature \cite{Walls}. We will show the connection between these phenomena through the thermal symmetry. We conclude with a discussion of our results.

\section{Thermal Symmetry in Markovian Master Equation (MME)}
\label{Thermalsym}

Consider a harmonic oscillator interacting with a thermal reservoir.
When memory effects can be neglected, e.g. in the weak coupling
limit, the time evolution of the reduced density operator $\rhoh$
of the harmonic oscillator is governed by a Markovian master
equation (MME), $ \d \rhoh / \d t = K \rhoh$. The time evolution
operator is \cite{Weidlich65,Barsegov02}
\begin{align} \label{Khat}
           K \rhoh = (K_0  + K_d )\rhoh \,,
\end{align}
where
\begin{align} \label{Khat0}
           K_0\rhoh
                 &= -i \ome_0 [ \adg a ,\rhoh \,] \\
            K_d\rhoh
              &=   c_1 ( 2 a  \rhoh \adg - \adg a \rhoh  - \rhoh \adg a ) \no
        &\qquad    + c_2 ( 2 \adg \rhoh a - a \adg \rhoh - \rhoh a \adg )
        \label{Khatd}\,,
\end{align}
in which we have made the rotating-wave approximation for the
interaction \cite{Cohen} and $a$ and $\adg$ satisfy the usual
commutation relation. All the dissipative effects of the dynamics
are contained in $K_d$. This MME generates a completely positive
dynamical semigroup on the reduced dynamics \cite{Kossa}.

In \Eq{Khat0}, $\ome_0$ is the natural frequency of the harmonic
oscillator and $c_1$, $c_2$ in \Eq{Khatd} are coefficients given by
\begin{align} \label{c1c2}
      c_1&=\frac{\gam}{2}\bigg(b+\frac{1}{2} \bigg)\,,
      &   c_2 &=  \frac{\gam}{2} \bigg( b- \frac{1}{2} \bigg)\,.
\end{align}
In the weak coupling approximation, the damping constant $\gam$ is
given by \cite{Barsegov02}
\begin{align} \label{gam}
      \gam &= \int_{0}^{\infty} d \ome 2\pi \lam^2 |v(\ome)|^2 \del(\ome - \ome_0) = 2\pi \lam^2 |v(\ome_0)|^2 \,,
\end{align}
where $\lam$ denotes the coupling constant of the interaction and
$v(\ome)$ is a model dependent form factor. The delta function in
\Eq{gam} arises from the contribution of the resonance pole at
$\ome_0$. The thermal properties of the reservoir are encoded in the
parameter $b$. In the weak coupling limit, $b$ can be determined
from the following expression
\begin{align} \label{gamb}
      \gam b &= \int_{0}^{\infty} d \ome 2\pi \lam^2 |v(\ome)|^2 \tilde{b}(\ome) \del(\ome - \ome_0) \no
            &= 2\pi \lam^2 |v(\ome_0)|^2 \tilde{b}(\ome_0) \,,
\end{align}
where we obtain
\begin{align} \label{b}
             b = \tilde{b}(\ome_0) \,,
\end{align}
after using \Eq{gam}. Assuming that the reservoir is in thermal
equilibrium, the parameter $\tilde{b}(\ome)$ is given by
\begin{align} \label{bome}
             \tilde{b}(\ome) =\frac{1}{2}+ \frac{1}{e^{ \hbar \ome \bt}-1}
        =\frac{1}{2} \coth \left( \half \hbar \ome \bt \right)
        \,,
\end{align}
where $\bt=1/k_B T$, $k_B$ being Boltzmann's constant and $T$ being
the temperature of the reservoir. The physical range of $b$ that
corresponds to $0 \leq T < \infty$ is $\half \leq b < \infty$.

We will now formulate our discussion in terms of the Liouville space
by following the superoperator formulation in TFD \cite{Umezawa}. We
introduce the (super-)operators
\begin{align} \label{A}
        \A &\equiv a \times 1 ,  & \A^\dg &\equiv \adg \times 1 \,.
\end{align}
and their tilde conjugate operators \footnote{The idea of tilde
conjugation $\Aa$ in TFD was introduced by Prigogine et. al. in
non-equilibrium statistical theory \cite{Prigogine73}, where it is
referred to as adjoint conjugation $\A^a$.}
\begin{align} \label{Atilde}
        \Aa &\equiv 1 \times \adg , & \Aa^\dg &\equiv 1 \times a \,.
\end{align}
These operators act on the density operator from the left according
to, e.g., $\A \Aa \rhoh = \A (\rhoh \adg) = a \rhoh \adg$, etc.
Hermitian and tilde conjugations are idempotent operations
\footnote{We consider only the bosonic operators. If fermionic
operators are included, two successive tilde conjugations will
result in an extra phase $\sig$, i.e. $\widetilde{\big(\At\big)} =
\sigma A$, where $\sig =1 $ for bososnic operators and $-1$ for
fermionic operators.},
\begin{align}
        (A^\dg)^\dg&=A \,, & \widetilde{\big(\At\big)} &= A \,,
\end{align}
and they act on the product of two arbitrary operators $X$, $Y$ as
\begin{align}
            (\eta XY)^\dg &= \eta^* Y^\dg X^\dg \,, &
            \widetilde{\big(\eta XY\big)} &= \eta^* \widetilde{X} \widetilde{Y} \,,
\end{align}
where $\eta$ is a complex number and $*$ denotes complex conjugate.
The commutation relations of the operators are
\begin{align} \label{AAdg}
        [\A,\A^\dg]&=1\,, & [\At, \At^\dg]&=1  \,,
\end{align}
while all other commutators vanish. We also have the useful relation
\begin{align} \label{AAAA}
        [\A-\At^\dg, (\A-\At^\dg\,)^\dg \,]=0\,.
\end{align}

In terms of these operators, the uncoupled and dissipative
components of the collision (super-)operator become
\begin{align}   \label{K0AA}
        K_0 &= -i \ome_0 \,  \big(\A^\dg \A - \At^\dg \At \big) \no
            &= -i \frac{ \ome_0 }{2} \left[  \big(\A +\At^\dg \big)^\dg  \big(\A-\At^\dg \big)
           + \big(\A+\At^\dg \big) \big(\A-\At^\dg \big)^\dg      \right]
\end{align}
and \footnote{$K_0$ and $K_d$ are invariant under tilde or adjoint
conjugation \cite{Prigogine73}, i.e. $\widetilde{K}_0=K_0$ and
$\widetilde{K}_d=K_d$.}
\begin{align} \label{KVAA}
        K_d
        &= \frac{\gam}{2} \left[  \big(\A - \At^\dg \big)^\dg  \big(\A+\At^\dg \big)
            - \big(\A -\At^\dg \big) \big(\A+\At^\dg \big)^\dg  \right] \no
            & \qquad -b \gam   \big(\A-\At^\dg \big)^\dg \big(\A-\At^\dg \big)
            \,,
\end{align}
respectively.

Now consider the unitary operator
\begin{align} \label{U}
        U(\theta) =e^{i G \theta} \,,
\end{align}
with the Hermitian generator
\begin{align} \label{Jso11}
            G \equiv i \big( \A \At - \A^\dg \At^\dg \big) \,.
\end{align}
Under the action of $U$ by conjugation, the $A$ operators are
transformed into
\begin{subequations}  \label{A'}
\begin{align}
          \A'&= U\A U^\dg = \cosh \theta \A - \sinh \theta \At^\dg  \,,
          \\
          \At'^\dg&= U\At^\dg U^\dg = - \sinh \theta \A + \cosh \theta \At^\dg
          \,,
\end{align}
\end{subequations}
with similar expressions for their Hermitian conjugate partners.
\Eq{A'} is a Bogoliubov transformation which preserves the
commutation relations in \Eq{AAdg}. This transformation can be
represented as a hyperbolic rotation on the operator pair $(\A,
\At^\dg)$ by \cite{Ban93,Santana06}
\begin{align} \label{Aprime}
        \left(
          \begin{array}{c}
            \A' \\
            \At'^\dg \\
          \end{array}
        \right)
        = R^{-1}(\theta)
        \left(
          \begin{array}{c}
            \A \\
            \At^\dg \\
          \end{array}
        \right)   ,
\end{align}
where the rotation matrix is
\begin{align} \label{Rtheta}
        R(\theta) = \left(
            \begin{array}{cc}
              \cosh \theta  &  \sinh \theta  \\
               \sinh \theta & \cosh \theta  \\
            \end{array}   \right) .
\end{align}
The transformation on $(A^\dg, \At)$ is obtained by taking the
Hermitian conjugation of \Eqs{A'} or (\ref{Aprime}). It should be
stressed that $R(\theta)$ (or $U$) mixes the operators $\A$ and its
tilde conjugate $\At^\dg$, and hence it is a rotation in the
Liouville space that has no counterpart in the Hilbert space.

The effect of $U$ on the operators $K_0$ and $K_d$ can now be
deduced by using the following linear combinations of \Eqs{A'}
\begin{align} \label{exptheta}
            \A'\mp \At'^\dg = e^{\pm \theta} \big( \A \mp \At^\dg \, \big) \,.
\end{align}
The uncoupled component of the dynamics $K_0$ has exact symmetry
under $U$,
\begin{align} \label{K0'}
        K'_0(A,\At^\dg) = U K_0(A,\At^\dg) U^\dg &= K_0(A',\At'^\dg) \no
        &= K_0(A,\At^\dg)\,,
\end{align}
that is, $[U,K_0]=0$. This symmetry in $K_0$ leads to the
degeneracies in the spectrum of the operator $L_0$ ($ = K_0$)
mentioned in the introduction. For the dissipative component of the
dynamics $K_d$, the first line on the rhs of \Eq{KVAA} has exact
symmetry under $U$, but the second line, which is proportional to
$b$, transforms up to a similarity transformation (or remains form
invariant). Overall, we have
\begin{align}   \label{KVU}
        K'_d(\A,\At^\dg;b)&  =  K_d(\A',\At'^\dg;b)= K_d(\A,\At^\dg;b')
        \,,
\end{align}
where the transformed parameter $b'$ is related to the angle of the
rotation by
\begin{align} \label{thetab}
            \theta = \half \ln (b'/b)\,.
\end{align}
For a reference thermal configuration $b$, the physical values of
$\theta$ lie within the range $-(\ln 2b ) /2 \leq \theta < \infty$,
for $\half \leq b' < \infty$. As a consequence of \Eqs{K0'} and
\eqref{KVU}, the hyperbolic rotation connects the full collision
operators between reservoirs of different temperatures up to a
similarity transformation
\begin{align}   \label{KAAb}
            K(b')   = U(\theta) K(b) U^\dg(\theta)  \,.
\end{align}
We note that in Ref.~\cite{Brink01}, the term ``Liouville symmetry"
refers to the existence of a symmetry $S$ that commutes with the
Liouville operator $\L$ (or the collision operator $K$), i.e.
$[L,S]=0$. For our system, while the thermal symmetry $\U(\theta)$
does not commute with $K$, it preserves the form of $K$ up to a
similarity transformation between different thermal states of the
system (\ref{KAAb}).

A quantum statistical system is described by a density operator
$\rhoh$ defined on the Hilbert space $\Hm$. As we formulate the
dynamics in the Liouville space $\Lm$, the system is now described
by a density state vector $|\rho\r$, which corresponds to $\rhoh$
\cite{Ban93,Petrosky97}. We have used the double ket notation to
emphasize the fact that the state vector lies in $\Lm$. Under the
thermal symmetry, the density state then transforms as
\begin{align}   \label{rho'}
        |\rho'(b)\r=U |\rho(b)\r = |\rho(b')\r
\end{align}
and the time evolution is now governed by $\d |\rho(b')\r/\d t =
K(b') |\rho(b')\r$.

The existence of the thermal symmetry is related to the degeneracies
in the eigenvalues of the collision operator $C(b)\equiv iK(b)$ in
the following way. Consider the non-Hermitian eigenvalue problem of
$C(b)$ \cite{Briegel93,Tay04,Morse},
\begin{align} \label{Cfmn}
            C(b) |\rho^\pm_{mn}(b)\r = z^\pm_{mn} |\rho^\pm_{mn}(b)\r \,,
\end{align}
which has complex eigenvalues denoted by
\begin{align} \label{zmn}
            z^\pm_{mn} &= \pm  n \ome_0 - i (m -n/2)\gam  \,, &  m\geq n \,,
\end{align}
where $m,n=0, 1, 2, 3, \ldots$, and the decay rate of the $(m,n)$
mode is $(m-n/2)\gam$. Under the action of $U$, \Eq{Cfmn} is written
\begin{align}   \label{degenerate}
    C(b') |\rho_{mn}^\pm(b')\r &= z^\pm_{mn} |\rho^\pm_{mn}(b')\r \,.
\end{align}
Since the complex eigenvalue $z^\pm_{mn}$ is independent of the
thermal parameter $b$, degeneracy in $z^\pm_{mn}$ can occur for
different thermal states of the system. This degeneracy in the
eigenvalues of the collision operator $K$ is a consequence of the
thermal symmetry.

The transformation $U$ can be interpreted as generating a continuous
change in the value of $T$ (or $\hbar$). For a transformation of the
system from a lower temperature to a higher one, we have $b'>b$
(equivalently, we can view this as a decrease in the value of $\hbar
$) with a positive angle, $\theta
> 0$. The system thus moves away from the quantum limit, while a transformation in
the opposite direction with $\theta < 0$ drives the system closer to
the quantum limit. In the following discussion, we shall fix the
value of $\hbar$ and interpret the transformation as generating a
change in the temperature.

Under the $U$ transformation, a system that is initially
characterized by $b$ is mapped into a different thermal state
characterized by $b'$. Hence the term {\it thermal} symmetry.
Moreover, all finite temperature states are related to the $T=0$ ($b
= \half$) state through the transformation. Since the $T=0$ state is
a purely quantum state, this implies that for systems with a thermal
symmetry, such as \Eq{Khat}, thermal effects can be inferred from
the effect of vacuum fluctuation by means of the thermal symmetry.

\section{Thermal symmetry in Other MMEs}
\label{coord}

In the previous section, the thermal symmetry is presented in the
superoperator form in \Eq{Aprime}, which clearly illustrates that
the symmetry cannot be reduced to a symmetry in the Hilbert space.
Other aspects of the symmetry can be discussed more conveniently in
terms of the coordinate basis representation. For instance, the
existence of a thermal symmetry in the classical Fokker-Planck
equation becomes obvious when we compare it with the quantum MME in
the Wigner's representation. The coordinate basis also proves to be
more convenient when we extend our discussion to include other MMEs,
where they acquire more simple forms.

We shall make use of the dimensionless coordinate $x\equiv (M \ome_0
/\hbar)^{1/2}\, q$, where $M$ is the mass of the harmonic oscillator
and $q$ is the ordinary spatial coordinate with dimensions of
length. In the Liouville space, we then have the pair of coordinates
$(x,\xt)$ to describe the system, which will transform as a
hyperbolic rotation similar to $\A$, cf. \Eq{ybar} below. Writing
the $A$ operators in terms of the position and momentum operators of
the harmonic oscillator, we have
\begin{align} \label{axp}
        A &= \tfrac{1}{\sqrt{2}} (\xh + i \ph) \,, & A^\dg &= \tfrac{1}{\sqrt{2}} (\xh - i
        \ph)\,,
\end{align}
where $\ph=-i\d/\d \xh$ is the dimensionless momentum. With similar
expressions for the tilde operators, we find that the operators and
their tilde counterparts separately satisfy the canonical
commutation relations,
\begin{align} \label{can}
        [\xh,\ph\,]&=i \,, & [\xth,\pth\,]=i \,,
\end{align}
while all other commutators vanish.

With the definition
\begin{align}
             \rho (x, \xt)\equiv \l x;\xt|\rho\r \equiv \< x | \rhoh |\xt \> \,,
\end{align}
the $\A$ operators are represented by \cite{Barsegov02,Tay04}
\begin{subequations} \label{Ainx}
\begin{align}
        \l x;\xt|\A|\rho\r  &= \frac{1}{\sqrt{2}} \left(x+\frac{\d}{\d x}\right) \rho(x,\xt) \,,
        \\
         \l x;\xt|\A^\dg|\rho\r  &= \frac{1}{\sqrt{2}} \left(x-\frac{\d}{\d x} \right) \rho(x,\xt)\,,
\end{align}
\end{subequations}
in the coordinate basis, with similar expressions for the tilde operators.

From \Eq{Ainx}, the pair of coordinates $(x,\xt)$ transforms under
the thermal symmetry just like the pair of $(\A,\At^\dg)$ operators,
\begin{align}   \label{ybar}
        \left(
          \begin{array}{c}
           x' \\
           \xt' \\
          \end{array}
        \right)
       = R^{-1}( \theta )
        \left(
          \begin{array}{c}
            x \\
            \xt \\
          \end{array}
        \right)    \,.
\end{align}
We also have the transformation equations
\begin{subequations} \label{xxtheta}
\begin{align} \label{xxa}
        x' \pm \xt' &= e^{\mp \theta} \, ( x \pm \xt) \,, \\
                \frac{\d}{\d x'} \pm \frac{\d}{\d \xt'} &=  e^{\pm \theta} \,
                \left(\frac{\d}{\d x} \pm \frac{\d}{\d \xt} \right) \label{dxdxa} \,,
\end{align}
\end{subequations}
where $\exp{\theta} = \sqrt{b'/b}$. In Sec.\ref{Sp}, we show that
\Eqs{xxtheta} can also be generated by a symplectic transformation.

In terms of the coordinate basis, the generator of the unitary
transformation is
\begin{align} \label{JxQ}
           G&= i \left( x \frac{\d}{\d {\xt}} + \xt \frac{\d}{\d x} \right)
             \,.
\end{align}

\subsection{Thermal symmetry in Fokker-Planck equation}

As a preparation to illustrate the existence of thermal symmetry in
the classical Fokker-Planck equation \cite{FokkerPlanck}, we first
write the quantum MME in the coordinate basis before establishing
its connection with the classical case.

In the coordinate basis, the collision operator takes the form
\begin{align} \label{KQr}
   & K(x,\xt;b) =  K_0(x,\xt) + K_d(x,\xt;b) \,,
\end{align}
where
\begin{align} \label{Koxx}
        K_0(x,\xt) &= -i \frac{\ome_0}{2}  \left( -\frac{\d^2}{\d  x^2}
           + \frac{\d^2}{\d  \xt^2}   +x^2-\xt ^2\right)
           \,, \\
           K_d(x,\xt;b) &= \frac{\gam}{4} \bigg[\left(\frac{\d}{\d x} +\frac{\d}{\d{\xt}} \right)(x+\xt)
             \no
             &\qquad\qquad\qquad -(x-\xt)\left( \frac{\d}{\d x} - \frac{\d}{\d{{\xt}}}
           \right)\bigg] \no
                 &\qquad     + b \frac{\gam }{ 2}  \left[\left(\frac{\d}{\d x} +\frac{\d}{\d{{\xt}}} \right)^2
                 -(x-\xt)^2\right]   \,, \label{Kdxx}
\end{align}
are written in a manifestly form invariant manner under the
transformation in \Eqs{xxtheta}.

As an intermediate step, we introduce a new pair of coordinates,
\begin{align} \label{Qr}
              Q &\equiv \half (x+\xt) \,, & r &\equiv x-\xt
              \,,
\end{align}
where the $(r,Q)$ coordinate system is related to $(x,\xt)$
coordinate system by a counter clock-wise rotation of $45^\text{o}$.
\Eqs{xxtheta} then imply that $(r,Q)$ transform separately under the
thermal symmetry as scale transformations
\begin{align} \label{Qrprime}
            Q' &= e^{-\theta} Q\,, & r' &= e^{\theta} r \,,
\end{align}
where $\exp(\theta)=\sqrt{b'/b}$.

To go into the Wigner representation, another pair of coordinates,
$(P,Q)$, where $P$ is the Fourier conjugate variable of $r$, are
used to describe the dynamics. Both coordinates transform with the
same scaling factor under the thermal symmetry,
\begin{align} \label{PQ}
            Q' &= e^{-\theta} Q \,, & P' &= e^{-\theta} P\,.
\end{align}
The collision operator in Wigner's representation is given by
\begin{multline}   \label{KPQ}
        K(Q,P;b) = -\ome_0  \left(P \frac{\d }{\d  Q}
                        -Q \frac{\d }{ \d  P}\right)   \\
         + \frac{\gam }{ 2}\left( \frac{\d }{ \d Q}Q+ \frac{\d }{ \d P}P \right)
                + b \frac{\gam }{ 2} \left( \frac{\d^2 }{ \d  Q^2}+ \frac{\d^2 }{ \d  P^2}\right)
                       .
\end{multline}
As is evident from \Eq{KPQ}, the thermal parameter $b$ can be
absorbed simultaneously into both coordinates, which is effectively
a scaling transformation according to \Eq{PQ}. The existence of the
second derivative terms in both the $Q$ and $P$ coordinates implies
that a Gaussian wave packet would tend to a more uniform
distribution along both the $Q$ and $P$-axis as the temperature
increases \cite{Walls}.

In Wigner's representation, the collision operator has the same form
as the classical Fokker-Planck equation \cite{FokkerPlanck}
expressed in the classical phase space variables, except with $b$
replaced by its classical analog
\begin{align}   \label{bcl}
        b_{cl}=k_B T/\hbar \ome_0 \,.
\end{align}
Therefore, we conclude that thermal symmetry exists in the classical
system as well. We can then run the argument in the reverse
direction, starting from a given classical Fokker-Planck equation,
we can transform the classical phase space variables corresponding
to $(P,Q)$ into the classical analog of the $(x,\xt)$ coordinate
system. A thermal hyperbolic symmetry can then be defined on these
coordinates as well. Indeed, the coordinates $(x,\xt)$ in the
classical system correspond to the Bargmann-Segal representation
\cite{Bargmann}, which is the classical analog of the coherent state
representation in quantum theory \cite{coherent}.

Scale transformations of the form in \Eq{PQ} also apply in the
coherent state representation of the MMEs \cite{Agarwal}, where the
pair of coordinates $z=P-iQ$ and its complex conjugate $z^*$ are
used. A pair of ``action-angle" variables $(J,\al)$, where
\begin{align}
        J &= \half (P^2+Q^2)   ,  & \al &= \tan^{-1}(Q/P) \,,
\end{align}
can also be defined for the quantum system in parallel to the
classical action-angle variables (for example, see \cite{Tay04}).
Under the thermal symmetry, $J$ then scales as $\exp(-2\theta) J$,
while $\al$ is left unaffected.

\subsection{Thermal symmetry in Caldeira-Leggett equation}

In the previous discussion, we have focused on the thermal symmetry
of the MME in \Eq{Khat}. We will now extend our consideration to
other quantum MMEs. It is interesting to recognize that the
Caldeira-Leggett (CL) equation \cite{CL} for the high temperature
limit of a harmonic oscillator interacting with a thermal reservoir
possesses this thermal symmetry as well. Indeed, the collision
operator of the CL equation is given by
\begin{align} \label{CL}
     K_{cl}(x,\xt;b_{cl})  &=  K_0(x,\xt)
            - \gam_1  (x-\xt) \left(\frac{\d}{\d x}-\frac{\d}{\d \xt}
            \right)\no
            &\qquad - \gam_1 b_{cl}  (x-\xt)^2
                        ,
\end{align}
where $b_{cl}$ in \Eq{bcl} is the high temperature limit of $b$. For
\Eq{CL}, the second line (with coefficient $b_{cl}$) transforms
under the thermal symmetry with an overall multiplicative factor
$\exp(2\theta)=b'_{cl}/b_{cl}$, whereas the other terms are
invariant under the transformation. We therefore have the following
transformation law for $K_{cl}$,
\begin{align} \label{KclU}
        K_{cl}(x,\xt;b_{cl}) \xrightarrow{U(\theta)} K_{cl}(x,\xt;b'_{cl}) \,,
\end{align}
and the CL equation is form invariant under the thermal symmetry.

\subsection{MMEs under more general conditions}

In order to determine the thermal symmetry for an MME, it is crucial
that we are able to define for $R(\theta)$ one single angle that is
independent of the frequencies of the field $\ome$. This global
property of the transformation originates from the resonance effect
expressed by the delta function $\del(\ome - \ome_0 )$ on the rhs of
\Eq{gamb}. Due to the delta function, $b$ is restricted to the
natural frequency of the reduced system $\ome_0 $.

However, for MMEs derived under more general considerations (for
finite temperature and without making the rotating-wave
approximation) \cite{Unruh89,HPZ92,Tay06}, the restriction of $\ome$
to $\ome_0$ does not occur in some of the coefficients. Let us
consider the Markovian limit of the collision operator for the MMEs
in Refs.~\cite{Unruh89,HPZ92,Tay06}. They have the general form
\begin{multline}  \label{exactMME}
        K_{g}(x,\xt;{b_k}) =  K_0(x,\xt)
      - \gam_2 (x-\xt) \left(\frac{\d}{\d x}-\frac{\d}{\d \xt} \right)     \\
          -   2 \gam_2 b (x-\xt)^2
         + i  \Gamma  (x-\xt) \left(\frac{\d}{\d x}+\frac{\d}{\d \xt} \right)
          \,,
\end{multline}
where
\begin{align}  \label{Lam}
        \Gamma = P \int_0^\infty d\ome \frac{I(\ome)}{\ome^2 - \ome_0^2} \, \tilde{b}(\ome) \,,
\end{align}
is a  temperature dependent coefficient. The symbol $P$ denotes the
principal value of the integration and $I(\ome)$ denotes the
spectral density of the field. Unlike the coefficient $\gam_2 b$,
cf. \Eq{gamb}, all frequency modes contribute to $\Gamma$ due to the
absence of a delta function in the integrand.

The first 3 terms on the rhs of \Eq{exactMME} have a structure
similar to the CL equation and hence are form invariant under
$U(\theta)$. However, the $\Gamma$-term destroys the form invariance
of $K_g$ for the following reason. To have the $\Gamma$-term remain
form invariant, we require
\begin{align}  \label{b'ome}
        \tilde{b}'(\ome) &= e^{2\theta} \tilde{b}(\ome)
\end{align}
for all $\ome$, or equivalently,
\begin{align}  \label{cothome}
       \frac{\coth( \hbar \ome/2 k_B T')}{\coth( \hbar \ome/2 k_B T)}
       &= \frac{\coth( \hbar \ome_0/2 k_B T')}{\coth( \hbar \ome_0/2 k_B T)}  \,,
\end{align}
after using \Eqs{bome}, \eqref{b} and \eqref{thetab}. This equation
has no non-trivial solutions for $T'\geq 0 $ other than $T'=T$, for
all $\ome > 0$.

We might think of introducing a transformation $U[\theta(\ome)]$
with a frequency dependent angle, where
\begin{align}
    \theta(\ome) \equiv \half \ln [\tilde{b}'(\ome) / \tilde{b}(\ome)] \,,
\end{align}
in order to restore the form invariance of $K_{g}$. However, since
the field degrees of freedom have been traced out to obtain the
MMEs, it is impossible to have all the $U[\theta(\ome)]$ (with
different $\ome$) acting on the $(x,\xt)$ coordinates
simultaneously. Therefore, this construction is not feasible.

One might also consider modification of \Eqs{xxtheta} in order to
incorporate a more general transformation that might account for the
form invariance of the $\Gamma$-term. However, the requirements for
the invariance of the canonical commutation relations in \Eqs{can}
along with those for the form invariance of the first three terms on
the rhs of \Eq{exactMME} under the modified transformation are too
restrictive for this modification to be possible. For instance, if
we modify the transformation coefficient for the derivative operator
in \Eq{dxdxa} from $\exp(\pm \theta)$ to $\exp(\pm \phi)$, where
$\phi$ is an angle to be determined by the form invariance
requirement for $\Gamma$, while we keep \Eq{xxa} unchanged, we find
that the second term on the rhs of \Eq{exactMME} no longer remains
form invariant. Furthermore, the modified operators no longer
satisfy the canonical commutation relations in \Eqs{can}.

In conclusion, the thermal symmetry in \Eq{Aprime} is applicable to
MMEs of linear systems: (1) under the rotating-wave approximation
with \Eq{Khat}, (2) in the high temperature system with the CL
equation \eqref{CL}, as well as (3) in the classical Fokker-Planck
equation. The symmetry does not apply to MMEs under more general
considerations, for example without the rotating-wave approximation
and for finite temperature. An example is the HPZ equation
\eqref{exactMME} \cite{HPZ92}. For non-linear systems,
generalization to the Bogoliubov transformations in \Eq{Rtheta}
should be envisaged.

\section{  Relations with Other Transformations }

In this section we discuss the relation of the thermal symmetry to
other forms of transformations utilized in closely related fields.

\subsection{Thermal symmetry as element of $\text{Sp}(2n,\mathbb{R})$ }
\label{Sp}

Since the thermal symmetry preserves the commutation relations of
the $A$ operators (\ref{AAdg}), it also preserves the canonical
commutation relations for the position and momentum operators
(\ref{can}). We can write the commutation relations in a more
compact form by introducing a column vector
\begin{align} \label{Qv}
        \Xh =
        \left(
          \begin{array}{c}
            \xh \\
            \xth \\
            \ph  \\
            \pth
          \end{array}
        \right)  \,.
\end{align}
The commutation relations in \Eq{can} can then be expressed as
\begin{align} \label{XX}
        [\Xh_i,\Xh_j]&=i \Omega_{ij} \,,& i,j&=1,2,3,4 \,,
\end{align}
where $\Omega$ is a non-singular, anti-symmetric matrix,
\begin{align} \label{b11}
        \Omega =   \left(
          \begin{array}{cc}
            0_2 & 1_2 \\
            - 1_2 & 0_2
          \end{array}
        \right) \,,
\end{align}
in which $1_2$ and $0_2$ denote the $2\times 2$ identity and null
matrix respectively.

\Eqs{A'} and their Hermitian conjugates can then be written as the action of
an element of the symplectic group on $\Xh$,
\begin{align} \label{XSX}
        \Xh'=S^{-1} \Xh \,,
\end{align}
where $S \in \text{Sp}(4,\mathbb{R})$ (with $n=2$) is given by
\begin{align} \label{S}
        S= e^{\theta J} =   \left(
          \begin{array}{cc}
            R(\theta) & 0_2  \\
            0_2 & R^{-1}(\theta)
          \end{array}
        \right) \,.
\end{align}
We have previously introduced the rotation matrix $R(\theta)$ in
\Eq{Rtheta}. The symplectic transformation $S$ is generated by
\begin{align} \label{J}
        J=   \left(
          \begin{array}{cc}
            \widetilde{1}_2 & 0_2  \\
            0_2 & -\widetilde{1}_2
          \end{array}
        \right) \,,
\end{align}
where
\begin{align} \label{1tilde}
        \widetilde{1}_2 =   \left(
          \begin{array}{cc}
           0 & 1  \\
           1 & 0
          \end{array}
        \right) \,.
\end{align}
By construction, the symplectic group $S$ preserves $\Omega$
\begin{align} \label{SBS}
        S \Omega S^T = \Omega \,,
\end{align}
as expected, where $T$ denotes the transpose operation.

The unitary operator $\bar{U}$ that corresponds to $S$ acts on
$\Xh$ by conjugation such that
\begin{align} \label{UQU}
        \bar{U}(S) \Xh \bar{U}^\dg(S) \equiv \Xh' = S^{-1} \Xh
\end{align}
by \Eq{XSX}. It can be expressed as $\bar{U}(S)=\exp[i \theta
\bar{G}(J)]$, where $\bar{G}(J)$ is the Hermitian generator of
$\bar{U}$ that corresponds to $J$. It has the following expression
\cite{Sudarshan87},
\begin{align}
        \bar{G}(J)=\half \Xh^T \Omega J \Xh=-(\xh \pth +\xth \ph)
\end{align}
which is identical to the generator of $U(\theta)$ in \Eq{JxQ} when
expressed in the coordinate basis. The transformation laws in
\Eqs{xxtheta} then follow from \Eq{UQU}.

Notice that we started with a one dimensional system, in which the
pair of canonical operators $(\xh,\ph)$ is available for describing
the dynamics. We are able to introduce the symplectic transformation
$\text{Sp}(4,\mathbb{R})$ on the four components vector $\Xh$ only
because the pair of tilde operators $(\xth,\pth)$ are introduced
when we go into the Liouville space.

\subsection{Relation of the thermal symmetry with $\text{SU}(1,1)$}

There is a close relationship between $U(\theta)$ and the
$\text{SU}(1,1)$ group. Consider the Lie algebra su(1,1), which
consists of 3 operators \cite{Ban93}
\begin{subequations} \label{su11}
\begin{align}
        M_1 &= \frac{1}{2}(\A^\dg \At^\dg + A \At)\,,\quad
        M_2 = \frac{1}{2i}(\A^\dg \At^\dg - A \At)\,,\\
        M_0&=\frac{1}{2} (\A^\dg \A+ \At^\dg \At +1 )\,.
\end{align}
\end{subequations}
These operators satisfy the commutation relations
\begin{subequations}
\begin{align} \label{su11comm}
       [M_1,M_2]&=-iM_0\,, & [M_2,M_0] &= iM_1 \,,\\
       [M_0,M_1]&=iM_2 \,.
\end{align}
\end{subequations}
An arbitrary $\text{SU}(1,1)$ transformation can be
parameterized as
\begin{align} \label{Usu}
  \U(\vec{\theta}\,) \equiv e^{-i\vec{\theta}\cdot \vec{M}}
      = e^{-i\theta_1 M_1 -i \theta_2 M_2 -i\theta_0 M_0} \,,
\end{align}
with inverse $\U^{-1}(\theta)=\U^\dg(\theta)=U(-\theta)$. By
choosing the angles as follows,
\begin{align}
        \theta_1&=\theta_0=0\,, & \theta_2 &=-2\theta \,,
\end{align}
Eq.(\ref{Usu}) reduces to \Eq{U}, or
$            \U(\theta) = \exp(i G \theta) $,
where the Hermitian generator
\begin{align} \label{JK-K+}
            G \equiv 2 M_2 = i \big( \A \At - \A^\dg \At^\dg \big)
\end{align}
generates the Lorentz group SO(1,1) in the $(x,\xt)$-plane
\cite{Gilmore}.

\subsection{Comparison with other form of the Bogoliubov transformation}

Let us compare the Bogoliubov (canonical) transformation discussed
in Ref.~\cite{Ekert90} with the Bogoliubov transformation in
\Eqs{A'}. It was shown in Ref.~\cite{Ekert90} that MME with the
following dissipative components
\begin{align} \label{Kvirtual}
            \underline{K}_d = K_d + c_3 K_3 + c_3^* K_3^\dg \,,
\end{align}
where
\begin{align} \label{K3}
            K_3 \rhoh = 2 a\rhoh a -aa\rhoh -\rhoh aa \,,
\end{align}
can be recast in the standard form of the dissipative operator $K_d$
in \Eq{Khat} by an SU(1,1) transformation on $(a,\adg)$,
\begin{align} \label{aunder}
            \underline{a} = u a + v \adg \,.
\end{align}
Here, $u=\cosh \nu$ and $v=\exp(i\eta)\sinh \nu $ are coefficients
related to $c_1$, $c_2$ and $c_3$. Even though this transformation
reduces $\underline{K}_d$ to the standard form $K_d$, the uncoupled
component of the dynamics $K_0$ (which need not be \Eq{K0AA}) may no
longer retain its original form.

In terms of the $A$ operators, \Eq{aunder} becomes
\begin{align} \label{Aunder}
            \underline{A} = u A + v A^\dg \,,
\end{align}
which unlike the thermal hyperbolic rotation (\ref{Aprime}), does
not mix the $A$ with the tilde operators, $\At$ or $\At^\dg$. Hence
this rotation {\it can} be reduced to a transformation in the
Hilbert space. This shows that the SU(1,1) hyperbolic rotation
symmetry of \Eq{aunder} is different from the thermal hyperbolic
rotation symmetry considered in Section \ref{Thermalsym}.

\section{Other Aspects of the Thermal Symmetry}

\subsection{Changes in the shape of Gaussian wave packet with temperature}

When the temperature of the reservoir changes, the profile of the
density function $\rho(x,\xt)$ changes in response \cite{Walls}. If
the temperature increases, the magnitude of the off-diagonal
components of $\rho$ decrease, a process called decoherence
\cite{Zurek03}. At the same time, the diagonal components of $\rho$
stretch out to attain a more uniform distribution. The density
function responds in the opposite way when the temperature
decreases. With a Gaussian wave-packet as an example, we shall
relate these behaviors of $\rho$ through the thermal symmetry.

Consider the equilibrium solution
of the collision operator $C(b)$ in \Eq{Cfmn}
in terms of the $(r,Q)$ coordinates \eqref{Qr}
\cite{Agarwal,Walls,Briegel93,Tay04},
\begin{align} \label{feqQr}
            \rhot(Q,r;b) = \frac{1}{\sqrt{2 \pi b}} \,
            \exp\left(-Q/2b - b r^2/2\right)
            \,,
\end{align}
which is a Gaussian wave packet in both the $r$ and $Q$ coordinates.

We now evaluate the averaged first and second moment of $Q$ and $r$
along the $r=0$ and $Q=0$ sections of $\rhot$. Along the $r=0$
section of $\rhot$, the average value of the first moment of $Q$
vanishes,
\begin{align} \label{Qr0}
           \<Q\>_{r=0} \equiv \frac{\int^{\infty}_{-\infty} Q \rhot(Q,0,b) dQ
           }{
           \int^{\infty}_{-\infty}  \rhot(Q,0,b) dQ} = 0 \,,
\end{align}
where the notation $\<\cdots\>_{r=0}$ means that we are evaluating
the average along the $r=0$ section of $\rhot$. We also find that
$\<r\>_{Q=0}=0$ \footnote{Note that $\<\cdots\>_{Q=0} $ is not the
usual trace average of an operator along the diagonal elements of
the density matrix.}. The second moments can also be evaluated
readily to give
\begin{align} \label{Q2r20}
        \<Q^2\>_{r=0} &=b\,, & \< r^2 \>_{Q=0}&=1/b \,.
\end{align}
The dispersion of the wave packet $(\Delta Q,\Delta r)$ is defined
as
\begin{align}
        \Delta Q = \sqrt{\< Q^2 \>_{r=0} - \< Q \>_{r=0}^2} \,,
\end{align}
with a corresponding expression for $\Delta r$ defined for the $Q=0$
section of $\rhot$. Substituting \Eqs{Qr0} and \eqref{Q2r20} into
$\Delta Q$ and $\Delta r$, we obtain
\begin{align}
        \Delta Q  &= \sqrt{b}  \,, & \Delta r &= 1/\sqrt{b} \,.
\end{align}

When the temperature increases, we have $b'>b$ (or $\theta>0$), and
the density function is transformed into
\begin{align}
        \rhot'(Q,r;b)=\rhot(Q',r';b)=\rhot(Q,r;b') \,.
\end{align}
The dispersion of the transformed Gaussian wave packet can be
evaluated to give
\begin{subequations} \label{DelQr'}
\begin{align}
        (\Delta Q)'  &= \sqrt{b'} = e^\theta \Delta Q  \,, \\
        (\Delta r)' &= \frac{1}{\sqrt{b'}}=e^{-\theta} \Delta r
        \,,
\end{align}
\end{subequations}
\Eqs{DelQr'} show that when the temperature increases ($\theta > 0$), the Gaussian wave packet
stretches away from the origin along the direction parallel to the
$Q$-axis ($r=0$ section), whereas it contracts towards the origin
along the direction parallel to the $r$-axis ($Q=0$ section). The
extent of the stretch and contraction is governed by the angle
$\theta$, or equivalently, the amount of change in the temperature.
The density function behaves in the opposite
manner when the temperature decreases.

\subsection{Disconnected regions in coordinate space}

When acting on the coordinate basis, the hyperbolic rotation
$R(\theta)$ preserves the bilinear form $x^2-\xt^2$. Each point on
the $(x,\xt)$-plane belongs to a family of curves $x^2-\xt^2 =
\text{const} $ generated by $R(\theta)$ and these curves never
cross. Consequently, the $(x,\xt)$-plane can be divided into 3
disconnected regions, according to $|x|=|\xt|$, $|x|>|\xt|$ and
$|x|<|\xt|$. The fact that these regions are disconnected shows that
the hyperbolic rotation does not mix quantum correlations
(off-diagonal component $\< x|\rho|\xt\>$) with probability
(diagonal component $\< x|\rho|x\>$). For the one dimensional case
considered in this paper, the $|x|>|\xt|$ $(|x|<|\xt|)$ regions are
further divided into 2 disconnected pieces, depending on the value
of $\mathrm{sgn}(x)=\pm$ $(\mathrm{sgn}(\xt)= \pm)$. For higher
dimensions, besides an overall hyperbolic rotation dependent on the
only available thermal parameter $b$, there may also be ordinary
rotations among the coordinates ${\bf{x}} = (x_1,x_2, ...)$ or
${\bf{\xt}} = (\xt_1,\xt_2, ...)$, without mixing between ${\bf{x}}$
and ${\bf{\xt}}$.

In analogy with special relativity, we can equally parameterize the
rotation matrix $R$ in terms of a ``velocity"-like parameter
$\text{v}$. Indeed, let us define $\text{v}$ by the relations
\cite{Santana06,Tay04}
\begin{align}
            \cosh\theta \equiv 1/\sqrt{1-\text{v}^2}\,, \qquad \sinh \theta \equiv \text{v}/\sqrt{1-\text{v}^2}\,.
\end{align}
We find that $\text{v}$ takes the familiar form
\begin{align} \label{v}
            \text{v}=\tanh\theta = \frac{b-\half}{b+\half}=\exp (-\hbar \ome_0 \bt) \,,
\end{align}
which is the relative probability of finding the frequency mode
$\ome_0 $ at thermal equilibrium. It asymptotically approaches the
``light"-speed $\text{v} \rightarrow 1$ (or $ \theta \to \infty$),
when $T \rightarrow \infty$, or $\hbar \to 0$ in the classical
limit. Hence, the classical limit as well as the high temperature
limit is a singular limit similar to the speed of light limit in
relativistic theory.

\section{Conclusion}

We have shown the existence of a thermal symmetry in the reduced
dynamics of open quantum systems with non-unitary time evolution,
for instance, in the MME with \Eq{Khat}, in the Caldeira-Leggett
equation, as well as in the classical Fokker-Planck equation.
However, for systems considered under more general conditions, such
as finite temperature and in the absence of the rotating-wave
approximation, the thermal symmetry is no longer present. The
Hu-Paz-Zhang equation is an example of the MMEs for such systems.

The thermal symmetry is generated by a Bogoliubov transformation on
the Liouville space of the reduced density operator. This symmetry
gives rise to degeneracies in the complex energy eigenvalues of the
dissipative collision operator. The presence of this symmetry is due
to the formulation of the dynamics on the level of the Liouville
space; it does not exist on the level of the Hilbert space.

As an important consequence of the symmetry, different thermal
states of the system are connected, including absolute zero. Hence
for systems observing the thermal symmetry, the effects of the
thermal reservoir on the system can be inferred from the properties
of the system at absolute zero, which does not contain thermal
effects. From a different point of view, the effects of the symmetry
can be regarded as changing the value of Planck's constant for the
case of fixed temperature. The symmetry then establishes a
connection between the quantum and classical limit of the system.

When represented in the coordinate basis, the thermal symmetry takes
the form of a hyperbolic rotation on the dynamical variables of the
reduced system. The angle of rotation depends on the amount of
change in the temperature, and it governs the extent of stretch and
contraction of the density function along the diagonal and
off-diagonal directions in the position coordinate basis, or the
$(x,\xt)$-plane.

The thermal symmetry can be expressed in terms of a Bogoliubov
transformation because the reduced system is a simple harmonic
oscillator, i.e., a linear reduced system. Generalization of this
symmetry to a nonlinear reduced system would require a
generalization of the Bogoliubov transformation \cite{Petrosky04}.
The thermal symmetry also enables the construction of a set of
temperature dependent density states of a coupled oscillator. We
will present the discussion elsewhere.

\acknowledgments

We thank Professor E.C.G.~Sudarshan and Professor W.C.~Schieve for
interesting discussions. We thank the referees for their
constructive comments and suggestions and for bringing
Refs.~\cite{Brink01,Ekert90,Ban93} to our attention. We also thank
Dr.~S.~Garmon for proofreading the manuscript. We acknowledge the
Engineering Research Program of the Office of Basic Energy Sciences
at the U.S.~Department of Energy, Grant No DE-FG03-94ER14465 for
supporting this work. B.A.~Tay acknowledges the support of the
Ministry of Science, Technology and Innovation, Malaysia (MOSTI)
Postdoctoral Research Scheme (STI) and thanks Dr.~H.~Zainuddin from
the Department of Physics at the University Putra Malaysia and
Dr.~T.L.~Yoon from the School of Physics at the University of
Science Malaysia for hospitality.


\end{document}